\documentstyle[amstex,amssymb]{article}

\catcode`\@=11
\def\numberbysection{\@addtoreset{equation}{section}
        \def\theequation{\thesection.\arabic{equation}}}
\numberbysection
\input{tcilatex}

\begin{document}

\newlength{\lno} \lno1.5cm \newlength{\len} \len=\textwidth%
\addtolength{\len}{-\lno}

\setcounter{page}{0}

\baselineskip7mm \renewcommand{\thefootnote}{\fnsymbol{footnote}} \newpage %
\setcounter{page}{0}

\begin{titlepage}     
\vspace{0.5cm}
\begin{center}
{\Large\bf Gaudin magnet with impurity and its generalized Knizhnik-Zamolodchikov equation}\\
\vspace{1cm}
{\large  A. Lima-Santos \hspace{.5cm} and \hspace{.5cm} Wagner Utiel } \\
\vspace{1cm}
{\large \em Universidade Federal de S\~ao Carlos, Departamento de F\'{\i}sica \\
Caixa Postal 676, CEP 13569-905~~S\~ao Carlos, Brazil}\\
\end{center}
\vspace{1.2cm}

\begin{abstract}
This work is concerned with the formulation of the  boundary quantum inverse scattering method  
for the xxz Gaudin magnet coupled to boundary impurities with arbitrary exchange constants. 
The Gaudin magnet is diagonalized by taking a quasi-classical limit of the inhomogeneous lattice. 
Using the method proposed by Babujian, the integral representation for the solution of the 
Knizhnik-Zamolodchikov equation is explictly constructed and its rational limit discussed.

\end{abstract}
\vspace{2cm}
\begin{center}
\end{center}
\vfill
\begin{center}
\small{\today}
\end{center}
\end{titlepage}

\baselineskip6mm

\newpage

\section{Introduction}

Integrable quantum field theories with boundaries have been subject of
intense study during the past decades. The great interest in such theories
stems from the large number of potential applications in different areas in
physics, including open strings, boundary conformal field theory,
dissipative quantum phenomena and impurity problems.

The Gaudin magnet has its origins in \cite{GA} as a quantum integrable model
describing $N$ spin-$\frac{1}{2}$ particles with long-range interactions.
The Gaudin type models have direct applications in condensed matter physics.
They also have been used as a testing ground for ideas such as the
functional Bethe ansatz ({\small BA}) and general procedure of separation of
variables \cite{Skly1, Skly2, Skly3}.

The model proposed by Gaudin was later generalized by several authors \cite%
{Jurco, Hikami1, Skly4}. The spin-s \ {\small XXY} Gaudin model was solved
in \cite{Babujian1} by means of the off-shell algebraic {\small BA}.

The {\small XYZ} Gaudin model was constructed and solved in \cite{Skly2} and 
\cite{Babujian2} by means of the algebraic {\small BA method}. The boundary 
{\small XXY} spin-$\frac{1}{2}$ Gaudin magnet was investigated by Hikami 
\cite{Hikami2} and \ the Gaudin models based on the face-type elliptic
quantum groups and boundary elliptic quantum group, as well as, the boundary 
{\small XYZ} Gaudin models were studied in \cite{Gould} by means of the
boundary algebraic {\small BA} method. In \cite{Yang} the {\small XXZ}
Gaudin model was solved with generic integrable boundaries specified by
generic non-diagonal $K$-matrices.

The Knizhnik-Zamolodchikov ({\small KZ}) equations were first proposed as a
set of differential equations satisfied by correlation functions of the
Wess-Zumino-Witten models \cite{KZ}. The relations between the Gaudin
magnets and the {\small KZ} equations has been studied in many papers \cite%
{Babujian1, Hasegawa, Feigin, LS1, LS2, LS3}. In \cite{Hikami2}, Hikami gave
an integral representation for the solutions of the {\small KZ} equations by
using the results of the boundary Gaudin model.

In addition, the quantum impurity problem, which has been extensively
investigated with renormalizing group techniques \cite{Kane} and conformal
field theory \cite{Affleck}, is also very interesting in itself. Andrei and
Johannesson \cite{Andrei} first considered an impurity spin-s embedded in an
integrable spin-$\frac{1}{2}$ {\small XXX} chain with periodic boundary
conditions. Subsequently, Schlottmann et al \cite{Schlottmann} generalized
it to the arbitrary spin chain. The standard approach to dealing with the
impurity integrable problem is also the algebraic {\small BA }method. The
Hamiltonian of the impurity integrable spin chain can be constructed from
the inhomogeneous transfer matrix. The key point is to find some
inhomogeneous vertex matrices, which satisfy the same Yang-Baxter relation
of the homogeneous matrices, corresponding to impurity spins.

We note that Sklyanin \cite{Skly4}, Mezincescu and Nepomechie \cite%
{Mezincescu} have used a constant number $K$ matrix to construct their
model, where the $K$ matrix induces the boundary fields and boundary bound
states \cite{Ghoshal, Fring}. In \cite{Wang1, Wang2}, Wang and coworkers
first introduced the operator $K$ matrix to study the Kondo problem in
one-dimensional strongly correlated electron systems. In a previous paper 
\cite{Wang3}, the problem of an open spin-$\frac{1}{2}$ Heisenberg chain
coupled to two spin-$s$ impurities sited at the ends has been studied.
Following this idea, Shu Chen {\it at al} \cite{Shu} have considered de 
{\small XXZ} chain coupled to impurity spins with different coupling
constants on the boundary.

In this paper, we continue to study the {\small XXZ} chain in order to
establish a link between the Gaudin model and the impurity problem. In the
first part we construct the eigenstate of the Gaudin magnet with impurity by
taking a quasi-classical limit of the transfer matrix for the inhomogeneous
open spin chain. The Hamiltonian is given as a solution of the classical
Yang-Baxter equation. In the second part, the {\small KZ} equation is
studied in the context of the impurity problem; the integral solution of the 
{\small KZ} equation is obtained in terms of the Bethe eigenstate of the
Gaudin magnet with impurity.

This paper is organized as follows. In section $2$ the boundary algebraic
Bethe ansatz is reviewed. The transfer matrix for $su(2)$ spin-$\frac{1}{2}$ 
{\small XXZ} chain is construct in terms of the $R$-matrix and the operator $%
K$ matrix. In section $3$ we find the Hamiltonian of the Gaudin magnet with
impurity by taking a quasi-classical limit of the double row transfer matrix
with operator $K$ matrices and subsequently its spectrum is obtained. In
section $4$ the off-shell Bethe ansatz of Babujian is used to find the
explicit integral solution of the {\small KZ} equation. The section $5$ is
reserved to a summary and discussion.

\section{The quantum inverse scattering method}

It is well know that in an integrable problem the quantum $R$-matrix
satisfies the Yang-Baxter equation ({\small YBE}): 
\begin{equation}
R_{12}(u-v)R_{13}(u)R_{23}(v)=R_{23}(v)R_{13}(u)R_{12}(u-v)  \label{qism.1}
\end{equation}%
As usual, $R_{ij}$ means the matrix on $V^{(1)}\otimes V^{(2)}\otimes
V^{(3)} $acting on the $i${\it th }and \ $j${\it th} spaces and as an
identity on the other spaces. The variables $u$ and $v$ are called the
spectral parameters. As a solution of {\small \ }(\ref{qism.1}), we use the $%
R$-matrix for the six-vertex model defined as%
\begin{equation}
R(u,\eta )=\left( 
\begin{array}{cccc}
1 & 0 & 0 & 0 \\ 
0 & b(u,\eta ) & c(u,\eta ) & 0 \\ 
0 & c(u,\eta ) & b(u,\eta ) & 0 \\ 
0 & 0 & 0 & 1%
\end{array}%
\right)  \label{qism.2}
\end{equation}%
where%
\begin{equation}
b(u,\eta )=\frac{\sinh u}{\sinh (u+\eta )},\qquad c(u,\eta )=\frac{\sinh
\eta }{\sinh (u+\eta )}.  \label{qism.3}
\end{equation}%
Note that in addition to the spectral parameter $u$, we have a deformation
parameter $\eta $ which parametrizes the anisotropy. Moreover, this solution
has the following properties:%
\begin{eqnarray}
{\em regularity} &{\em :}&{\em \quad }R(u=0,\eta )=P  \nonumber \\
{\em quasi-classical\ condition} &{\em :}&{\em \quad }R(u,\eta =0)=1
\label{qism.4} \\
{\em unitarity} &{\em :}&{\em \quad }R(u,\eta )R(-u,\eta )=1  \nonumber
\end{eqnarray}%
where $P$ is the permutation operator: $P\left\vert \alpha \right\rangle
\otimes \left\vert \beta \right\rangle =\left\vert \beta \right\rangle
\otimes \left\vert \alpha \right\rangle $ and $t_{i}$ denotes the
transposition of the $i${\it th} space.

Let us define the monodromy matrix $T(u)$ for the inhomogeneous $N$-sites
spin chain, introducing the inhomogeneity in the lattice through the
parameter $z_{N}\in C,$ by%
\begin{equation}
T_{0}(u|z)=R_{0N}(u-z_{N})\cdots R_{02}(u-z_{2})R_{01}(u-z_{1})=\left( 
\begin{array}{cc}
A(u|z) & B(u|z) \\ 
C(u|z) & D(u|z)%
\end{array}%
\right)  \label{qism.5}
\end{equation}%
Here the operator matrix elements act on the full Hilbert space $V^{\otimes
N}$. Due to the additive property of the spectral parameter, the {\small YBE}
also holds for the inhomogeneous lattice and we have%
\begin{equation}
R_{12}(u-v)\left[ T(u|z)\otimes T(v|z)\right] =\left[ T(v|z)\otimes T(u|z)%
\right] R_{12}(u-v)  \label{qism.6}
\end{equation}%
Equation (\ref{qism.6}) gives the fundamental algebraic structure for the 
{\small QISM} and gives us the commutation relations between the operators $%
A(u|z)$,$B(u|z)$,$C(u|z)$ and $D(u|z)$%
\begin{eqnarray}
\lbrack B(u|z),B(v|z)] &=&0  \nonumber \\
A(u|z)B(u|z) &=&\frac{1}{b(v-u)}B(v|z)A(u|z)-\frac{c(v-u)}{b(v-u)}%
B(u|z)A(v|z)  \nonumber \\
D(u|z)B(v|z) &=&\frac{1}{b(u-v)}B(v|z)D(u|z)-\frac{c(u-v)}{b(u-v)}%
B(u|z)D(v|z)  \nonumber \\
\lbrack B(u|z),C(v|z)] &=&\frac{c(u-v)}{b(u-v)}\left(
D(v|z)A(u|z)-D(u|z)A(v|z)\right)  \label{qism.7}
\end{eqnarray}%
To construct the eigenstate of our system, we can use the reference state $%
\left\vert 0\right\rangle $%
\begin{equation}
\left\vert 0\right\rangle =\left( 
\begin{array}{c}
1 \\ 
0%
\end{array}%
\right) _{1}\otimes \cdots \otimes \left( 
\begin{array}{c}
1 \\ 
0%
\end{array}%
\right) _{N}  \label{qism.8}
\end{equation}%
It is easy to check that this state is the eigenstate of the operators $%
A(u|z)$ and $D(u|z)$, and annihilated by $C(u|z)$:%
\begin{equation}
A(u|z)\left\vert 0\right\rangle =\left\vert 0\right\rangle ,\quad
D(u|z)\left\vert 0\right\rangle =\prod\limits_{k=1}^{N}\frac{\sinh (u-z_{k})%
}{\sinh (u-z_{k}+\eta )}\left\vert 0\right\rangle ,\quad C(u|z)\left\vert
0\right\rangle =0  \label{qism.9}
\end{equation}%
We note that the Bethe state, a sum of spin waves \cite{Faddeev}, can be
generated by operators $B(u|z)$ acting on the reference state $\left\vert
0\right\rangle .$

In order to construct an integrable open chain with boundary impurities, it
is necessary to introduce the reflection matrices $K^{-}(u)$ and $K^{+}(u)$
which satisfy the following reflecting equations \cite{Skly4}:%
\begin{equation}
R_{12}(u-v)K_{1}^{-}(u)R_{12}(u+v)K_{2}^{-}(v)=K_{2}^{-}(v)R_{12}(u+v)K_{1}^{-}(u)R_{12}(u-v)
\label{qism.10}
\end{equation}%
\begin{equation}
R_{12}(-u+v)K_{1}^{+t_{1}}(u)R_{12}(-u-v-2\eta
)K_{2}^{+t_{2}}(v)=K_{2}^{+t_{2}}(v)R_{12}(-u-v-2\eta
)K_{1}^{+t_{1}}(u)R_{12}(-u+v)  \label{qism.11}
\end{equation}%
The solution $K_{1}^{\pm }=K^{\pm }\otimes I$ and $K_{2}^{\pm }=I\otimes
K^{\pm }$ are the simplest reflection matrices which satisfies the
reflecting equations. The inhomogeneous transfer matrix $t(u)$ is defined as%
\begin{equation}
t(u|z)=Tr_{0}(K_{0}^{+}(u)T_{0}(u|z)K_{0}^{-}(u)T_{0}^{-1}(-u|z))
\label{qism.12}
\end{equation}%
and forms a one-parameter commutative family%
\begin{equation}
\left[ t(u|z),t(v|z)\right] =0  \label{qism.13}
\end{equation}%
where the monodromy matrix $T(u|z)$ is given by (\ref{qism.5}) and by virtue
of the unitarity property of our $R$-matrix (\ref{qism.4}), it follows the
expression for the reflected monodromy matrix 
\begin{equation}
T_{0}^{-1}(-u|z)=R_{01}(u+z_{1})R_{02}(u+z_{2})\cdots R_{0N}(u+z_{N})=\left( 
\begin{array}{cc}
A(u|-z) & B(u|-z) \\ 
C(u|-z) & D(u|-z)%
\end{array}%
\right)   \label{qism.14}
\end{equation}

It was proved in \cite{Shu} that if $\tau $ obeys the Yang relation%
\begin{equation}
R_{12}(u-v)\tau _{1}(u)\tau _{2}(v)=\tau _{2}(u)\tau _{1}(v)R_{12}(u-v)
\label{qism.15}
\end{equation}%
then%
\begin{equation}
K^{-}(u)=\tau (u+c)\tau ^{-1}(-u+c)  \label{qism.16}
\end{equation}%
also obeys the reflecting equation (\ref{qism.10}) and $c$ is a constant.

Therefore we can construct our reflection matrix as%
\begin{equation}
K_{0}^{-}(u)=R_{0L}(u+c_{L})R_{0L}^{-1}(-u+c_{L})=R_{0L}(u+c_{L})R_{0L}(u-c_{L})
\label{qism.17}
\end{equation}%
where $c_{L}$ is a constant decided by the left boundary. This construction
give us an operator $K$-matrix instead of a constant numerical matrix, where
it is identified as impurity and is not a pure reflection.

In order to obtain the integrable Hamiltonian \ in the open chain we define%
\begin{equation}
U_{a}(u|z)=K_{0}^{+}(u){\cal M}_{0}(u|z)K_{0}^{-}(u){\cal M}_{0}^{-1}(-u|z)
\label{qism.18}
\end{equation}%
where $K_{0}^{+}(u)=I$ and $K_{0}^{-}(u)$ is given by (\ref{qism.17}), and $%
{\cal M}_{0}(u|z),$ ${\cal M}_{0}^{-1}(-u|z)$ are defined as 
\begin{equation}
{\cal M}_{0}(u|z)=R_{0R}(u+c_{R})T_{0}(u|z),\qquad {\cal M}%
_{0}^{-1}(u|z)=T_{0}^{-1}(-u|z)R_{0R}(u-c_{R})  \label{qism.19}
\end{equation}%
Here $c_{R}$ is a constant decided by the right boundary. Now one can prove
that $U(u)$ satisfies the reflecting equation%
\begin{equation}
R_{12}(u-v)U_{1}(u)R_{12}(u+v)U_{2}(v)=U_{2}(v)R_{12}(u+v)U_{1}(u)R_{12}(u-v).
\label{qism.20}
\end{equation}

In order to derive the corresponding Hamiltonian we first recover the
homogeneous case taking $z_{i}=0,i=1,..N$ and the new transfer matrix%
\begin{equation}
X(u)=Tr(K_{0}^{+}(u){\cal M}_{0}(u)K_{0}^{-}(u){\cal M}_{0}^{-1}(-u))
\label{qism.21}
\end{equation}

The Hamiltonian can be obtained by

\begin{eqnarray}
H &=&\frac{d}{du}\ln X(u)|_{u=0}=\sum_{j=1}^{N-1}\left( \sigma
_{j}^{-}\sigma _{j+1}^{+}+\sigma _{j}^{+}\sigma _{j+1}^{-}+\cosh \eta \
\sigma _{j}^{z}\sigma _{j+1}^{z}\right)  \nonumber \\
&&+\cosh c_{L}\left( \sigma _{1}^{-}\sigma _{L}^{+}+\sigma _{1}^{+}\sigma
_{L}^{-}\right) +\cosh \eta \ \sigma _{1}^{z}\sigma _{L}^{z}  \nonumber \\
&&+\cosh c_{R}\left( \sigma _{N}^{-}\sigma _{R}^{+}+\sigma _{N}^{+}\sigma
_{R}^{-}\right) +\cosh \eta \ \sigma _{N}^{z}\sigma _{R}^{z}  \label{qism.22}
\end{eqnarray}

This Hamiltonian describes, beside the bulk, the interaction of the
particles with the impurities in the magnetic system. The contribution of
the right and left impurities is explicitly in the form of the Hamiltonian.

In the following we will use the algebraic Bethe ansatz \cite{Skly4} to
solve the spectrum of $X(u)$ for the inhomogeneous $N$-site spin chain.

The double row monodromy matrix can be written as%
\begin{equation}
U_{a}(u|z)=\left( 
\begin{array}{cc}
{\cal A}(u|z) & {\cal B}(u|z) \\ 
{\cal C}(u|z) & {\cal D}(u|z)%
\end{array}%
\right)  \label{qism.23}
\end{equation}%
Note that the operators ${\cal A}$,${\cal B}$,${\cal C}$ and ${\cal D}$ act
on the Hilbert space $V^{\otimes N}$. The boundary Yang-Baxter relation (\ref%
{qism.10}) can be rewritten in terms of the operator matrix elements of $%
U_{a}(u|z).$For convenience we define the operator%
\begin{equation}
\overset{\wedge }{{\cal D}}(u|z)=\sinh (2u+\eta ){\cal D}(u|z)-\sinh \eta \ 
{\cal A}(u|z).  \label{qism.24}
\end{equation}%
then we will have the following commutation relations%
\begin{equation}
\left[ {\cal B}(u|z),{\cal B}(v|z)\right] =0  \label{qism.25}
\end{equation}%
\begin{eqnarray}
{\cal A}(u|z){\cal B}(v|z) &=&\frac{\sinh (u+v)\sinh (u-v-\eta )}{\sinh
(u+v+\eta )\sinh (u-v)}{\cal B}(v|z){\cal A}(u|z)  \nonumber \\
&&+\frac{\sinh (2v)\sinh \eta }{\sinh (2v+\eta )\sinh (u-v)}{\cal B}(u|z)%
{\cal A}(v|z)  \nonumber \\
&&-\frac{\sinh \eta }{\sinh (u+v+\eta )\sinh (2v+\eta )}{\cal B}(u|z)\overset%
{\wedge }{{\cal D}}(v|z)  \label{qism.26}
\end{eqnarray}%
\begin{eqnarray}
\overset{\wedge }{{\cal D}}(u|z){\cal B}(v|z) &=&\frac{\sinh (u-v-\eta
)\sinh (u+v+2\eta )}{\sinh (u-v)\sinh (u+v+\eta )}{\cal B}(v|z)\overset{%
\wedge }{{\cal D}}(u|z)  \nonumber \\
&&-\frac{\sinh (2u+2\eta )\sinh \eta }{\sinh (2v+\eta )\sinh (u-v)}{\cal B}%
(u|z)\overset{\wedge }{{\cal D}}(v|z)  \nonumber \\
&&+\frac{\sinh \eta \sinh (2v)\sinh (2u+2\eta )}{\sinh (u+v+\eta )\sinh
(2v+\eta )}{\cal B}(u|z){\cal A}(v|z)  \label{qism.27}
\end{eqnarray}

The transfer matrix $X(u)$ can be expressed as%
\begin{eqnarray}
X(u|z) &=&{\cal A}(u|z)+{\cal D}(u|z)  \nonumber \\
&=&\frac{\sinh (2u+\eta )+\sinh \eta }{\sinh (2u+\eta )}{\cal A}(u|z)+\frac{1%
}{\sinh (2u+\eta )}\overset{\wedge }{{\cal D}}(u|z)  \label{qism.28}
\end{eqnarray}%
Now we define a reference state by including the two boundary impurities
sites%
\begin{equation}
\left\vert 0\right\rangle =\left( 
\begin{array}{c}
1 \\ 
0%
\end{array}%
\right) _{L}\otimes \left( 
\begin{array}{c}
1 \\ 
0%
\end{array}%
\right) _{1}\otimes \cdots \otimes \left( 
\begin{array}{c}
1 \\ 
0%
\end{array}%
\right) _{N}\otimes \left( 
\begin{array}{c}
1 \\ 
0%
\end{array}%
\right) _{R}  \label{qism.29}
\end{equation}%
This reference state is an eigenstate of ${\cal A}(u|z)$ and $\overset{%
\wedge }{{\cal D}}(u|z)$ and annihilated by ${\cal C}(u|z)$%
\begin{equation}
{\cal A}(u|z)\left\vert 0\right\rangle =\left\vert 0\right\rangle ,\quad 
\overset{\wedge }{{\cal D}}(u|z)\left\vert 0\right\rangle =\overset{\wedge }{%
d}(u|z)\left\vert 0\right\rangle ,\quad {\cal C}(u|z)\left\vert
0\right\rangle =0  \label{qism.30}
\end{equation}%
where%
\begin{eqnarray}
\overset{\wedge }{d}(u|z) &=&2\sinh (u)\cosh (u+\eta )\prod\limits_{a=L,R}%
\frac{\sinh (u+c_{a})\sinh (u-c_{a})}{\sinh (u+c_{a}+\eta )\sinh
(u-c_{a}+\eta )}  \nonumber \\
&&\times \left[ \prod\limits_{k=1}^{N}\frac{\sinh (u-z_{k})\sinh (u+z_{k})}{%
\sinh (u-z_{k}+\eta )\sinh (u+z_{k}+\eta )}\right]  \label{qism.31}
\end{eqnarray}

The eigenstate of $X(u)$ with $M$ spins down is given by the Bethe state%
\begin{equation}
\Psi (\{u\})=\prod\limits_{a=1}^{M}{\cal B}(u_{a}|z)\left\vert 0\right\rangle
\label{qism.32}
\end{equation}%
Using the commutation relations between operators ${\cal A}(u|z)$, ${\cal B}%
(u|z)$ and $\overset{\wedge }{{\cal D}}(u|z)$, we obtain%
\begin{equation}
X(u)\Psi (\{u\})=\Lambda (u;\{u\}|z)\Psi (\{u\})+\sum_{a}{\cal F}_{a}\Psi
^{a}(\{u\})  \label{qism.33}
\end{equation}%
where%
\begin{eqnarray}
\Lambda (u;\{u\}|z) &=&\frac{\sinh (2u+\eta )+\sinh \eta }{\sinh (2u+\eta )}%
\prod\limits_{a=1}^{M}\frac{\sinh (u-u_{a}-\eta )\sinh (u+u_{a})}{\sinh
(u-u_{a})\sinh (u+u_{a}+\eta )}  \nonumber \\
&&+\frac{\overset{\wedge }{d}(u|z)}{\sinh (2u+\eta )}\prod\limits_{a=1}^{M}%
\frac{\sinh (u-u_{a}+\eta )\sinh (u+u_{a}+2\eta )}{\sinh (u-u_{a})\sinh
(u+u_{a}+\eta )}  \label{qism.34}
\end{eqnarray}%
\begin{eqnarray}
{\cal F}_{a} &=&\left\{ \left[ \prod\limits_{b\neq a}^{M}\frac{\sinh
(u_{a}-u_{b}-\eta )\sinh (u_{a}+u_{b})}{\sinh (u_{a}-u_{b})\sinh
(u_{a}+u_{b}+\eta )}\right] \sinh (2u_{a})\cosh (u_{a})\right.  \nonumber \\
&&-\left. \left[ \prod\limits_{b\neq a}^{M}\frac{\sinh (u_{a}-u_{b}+\eta
)\sinh (u_{a}+u_{b}+2\eta )}{\sinh (u_{a}-u_{b})\sinh (u_{a}+u_{b}+\eta )}%
\right] \cosh (u_{a}+\eta )\overset{\wedge }{d}(u_{a}|z)\right\}  \nonumber
\\
&&\times \frac{2\sinh (u+\eta )\sinh \eta }{\sinh (u-u_{a})\sinh
(2u_{a}+\eta )\sinh (u+u_{a}+\eta )}  \label{qism.35}
\end{eqnarray}%
and%
\begin{equation}
\Psi ^{a}(\{u\})={\cal B}(u|z)\prod\limits_{b\neq a}^{M}{\cal B}%
(u_{b}|z)\left\vert 0\right\rangle  \label{qism.36}
\end{equation}

The relation (\ref{qism.33}) shows that the Bethe state $\Psi (\{u\})$ is an
eigenstate of the transfer matrix $X(u)$ under the condition ${\cal F}_{a}=0$%
, $a=1,...,M$, i.e.%
\begin{eqnarray}
&&\prod\limits_{b\neq a}^{N}\frac{\sinh (u_{a}-z_{b}+\eta )}{\sinh
(u_{a}-z_{b})}\frac{\sinh (u_{a}+z_{b}+\eta )}{\sinh (u_{a}+z_{b})}=\frac{%
\cosh ^{2}(u_{a}+\eta )}{\cosh ^{2}u_{a}}  \nonumber \\
&&\times \prod\limits_{j=L,R}\frac{\sinh (u_{a}+c_{j})\sinh (u_{a}-c_{j})}{%
\sinh (u_{a}+c_{j}+\eta )\sinh (u_{a}-c_{j}+\eta )}\prod\limits_{b\neq a}^{M}%
\frac{\sinh (u_{a}-u_{b}+\eta )\sinh (u_{a}+u_{b}+2\eta )}{\sinh
(u_{a}-u_{b}-\eta )\sinh (u_{a}+u_{b})}  \label{qism.37}
\end{eqnarray}%
This equation with $z_{k}=0$ corresponds to the Bethe ansatz equation for 
{\small XXZ} spin chain with impurities derived in \cite{Shu}.

\section{The Gaudin magnet}

We will show how the Gaudin magnet with impurity can be derived from the
identity (\ref{qism.33}). The Gaudin magnet can be obtained by taking the
quasi-classical limit $\eta \rightarrow 0$ of the transfer matrix for the
inhomogeneous spin chain \cite{GA}. This fact indicates that the Hamiltonian
is written in terms of the solution of the classical {\small YBE}.

Due to the quasi-classical condition, we have the power series expansion
around the point $\eta =0$ for each term in (\ref{qism.33},\ref{qism.34}).

\begin{equation}
X(u=z_{j})=-\eta +\eta ^{2}H_{j}+o(\eta ^{3}),\qquad \Lambda (u=z_{j})=-\eta
+\eta ^{2}E_{j}+o(\eta ^{3}),  \label{gm.1}
\end{equation}%
and from (\ref{qism.35}) we have%
\begin{equation}
{\cal F}_{a}=-\eta ^{2}\frac{2\sinh z_{j}\cosh v_{a}}{\sinh
(z_{j}-v_{a})\sinh (z_{j}+v_{a})}f_{a}+o(\eta ^{3})  \label{gm1a}
\end{equation}%
where we have the Hamiltonian $H_{j}$: 
\begin{eqnarray}
H_{j} &:&=\sum_{k=1}^{N}\frac{1}{\sinh (z_{j}+z_{k})}\left\{ \sigma
_{j}^{-}\sigma _{k}^{+}+\sigma _{j}^{+}\sigma _{k}^{-}+\frac{1}{2}\cosh
(z_{j}+z_{k})\left( \sigma _{j}^{z}\sigma _{k}^{z}+1\right) \right\} 
\nonumber \\
&&+\sum_{k\neq j,k=1}^{N}\frac{1}{\sinh (z_{j}-z_{k})}\left\{ \sigma
_{j}^{-}\sigma _{k}^{+}+\sigma _{j}^{+}\sigma _{k}^{-}+\frac{1}{2}\cosh
(z_{j}-z_{k})\left( \sigma _{j}^{z}\sigma _{k}^{z}+1\right) \right\} 
\nonumber \\
&&+\frac{2\sinh z_{j}}{\sinh (z_{j}-c_{L})\sinh (z_{j}+c_{L})}\left\{ \cosh
c_{L}\left( \sigma _{j}^{-}\sigma _{L}^{+}+\sigma _{j}^{+}\sigma
_{L}^{-}\right) +\frac{1}{2}\cosh (z_{j})\left( \sigma _{j}^{z}\sigma
_{L}^{z}+1\right) \right\}  \nonumber \\
&&+\frac{2\sinh z_{j}}{\sinh (z_{j}-c_{R})\sinh (z_{j}+c_{R})}\left\{ \cosh
c_{R}\left( \sigma _{R}^{-}\sigma _{j}^{+}+\sigma _{R}^{+}\sigma
_{j}^{-}\right) +\frac{1}{2}\cosh (z_{j})\left( \sigma _{R}^{z}\sigma
_{j}^{z}+1\right) \right\} ,  \nonumber \\
&&  \label{gm.2}
\end{eqnarray}%
the energy $E_{j}$: 
\begin{eqnarray}
E_{j} &=&\frac{1}{\sinh (2z_{j})}+\sum_{k=L,R}\left[ \coth
(z_{j}-c_{k})+\coth (z_{j}+c_{k}\right]  \nonumber \\
&&-\sum_{a=1}^{M}\left[ \coth (z_{j}-v_{a})+\coth (z_{j}+v_{a})\right]
\label{gm.3}
\end{eqnarray}%
and the unwanted factor $f_{a}$: 
\begin{eqnarray}
f_{a} &=&2\tanh v_{a}-\sum_{k=L,R}\left[ \coth (v_{a}-c_{k})+\coth
(v_{a}+c_{k})\right]  \nonumber \\
&&+2\sum\begin{Sb} b=1  \\ b\neq a  \end{Sb}  ^{M}\left[ \coth
(v_{a}-v_{b})+\coth (v_{a}+v_{b})\right] -\sum_{k=1}^{N}\left[ \coth
(v_{a}-z_{k})+\coth (v_{a}+z_{k})\right]  \label{gm.4}
\end{eqnarray}%
Note that we have used the notation $(u=z_{j})$ to mean the residue at $%
u=z_{j}$. The integrability of these Hamiltonians follow from their
commutativity%
\begin{equation}
\left[ H_{j},H_{k}\right] =0,\qquad j=1....,N  \label{gm.5}
\end{equation}%
which is obtained from the commutativity of the transfer matrix.

The Bethe states $\Psi $ (\ref{qism.32}) and $\Psi _{a\text{ }}$(\ref%
{qism.36}) have the following expansions%
\begin{equation}
\Psi (v)=\eta ^{M}\phi +o(\eta ^{M+1}),\qquad \Psi ^{a}(z_{j})=\eta
^{M-1}\sigma _{j}^{-}\phi ^{a}+o(\eta ^{M})  \label{gm.6}
\end{equation}%
with%
\begin{equation}
\phi =\prod\limits_{a=1}^{M}\left\{ {\cal Z}_{a}+\sum_{k=1}^{N}\left( \frac{1%
}{\sinh (v_{a}-z_{k})}+\frac{1}{\sinh (v_{a}+z_{k})}\right) \sigma
_{k}^{-}\right\} \left\vert 0\right\rangle  \label{gm.7}
\end{equation}%
and%
\begin{equation}
\phi ^{a}=\prod\limits\begin{Sb} b=1  \\ b\neq a  \end{Sb}  ^{M}\left\{ 
{\cal Z}_{b}+\sum_{k=1}^{N}\left( \frac{1}{\sinh (v_{b}-z_{k})}+\frac{1}{%
\sinh (v_{b}+z_{k})}\right) \sigma _{k}^{-}\right\} \left\vert 0\right\rangle
\label{gm.8}
\end{equation}%
where ${\cal Z}_{a}$ are the impurity contributions 
\begin{equation}
{\cal Z}_{a}=\left( \frac{1}{\sinh (v_{a}-c_{L})}+\frac{1}{\sinh
(v_{a}+c_{L})}\right) \sigma _{L}^{-}+\left( \frac{1}{\sinh (v_{a}-c_{R})}+%
\frac{1}{\sinh (v_{a}+c_{R})}\right) \sigma _{R}^{-}  \label{gm.9}
\end{equation}

When we combine the terms proportional to $\eta ^{M+1}$ in (\ref{qism.33}),
we obtain the so-called off-shell Bethe ansatz equation%
\begin{equation}
H_{j}\phi =E_{j}\phi +\sum_{a=1}^{M}\frac{2\sinh z_{j}\cosh v_{a}}{\sinh
(z_{j}-v_{a})\sinh (z_{j}+v_{a})}f_{a}\ \sigma _{j}^{-}\phi ^{a},\qquad
j=1,...,N  \label{gm.10}
\end{equation}%
This equation suggests the Bethe state $\phi $ (\ref{gm.7}) is an eigenstate
of the Gaudin's Hamiltonian $H_{j}$, if the set of rapidities $\{v_{a}\}$
satisfy $f_{a}=0$ ($a=1,...,M$), i.e. the Bethe equations:%
\begin{eqnarray}
&&\sum_{k=1}^{N}\left[ \coth (v_{a}-z_{k})+\coth (v_{a}+z_{k})\right] 
\nonumber \\
&=&2\tanh v_{a}-\sum_{k=L,R}\left[ \coth (v_{a}-c_{k})+\coth (v_{a}+c_{k})%
\right] +2\sum\begin{Sb} b=1  \\ b\neq a  \end{Sb}  ^{M}\left[ \coth
(v_{a}-v_{b})+\coth (v_{a}+v_{b})\right]  \nonumber \\
&&  \label{gm.11}
\end{eqnarray}

We have derived the eigenstate and the energy of the {\small XXZ}-type
Gaudin magnet coupled with impurity spin with different coupling constants $%
c_{L\text{ }}$and $c_{R}$ on the boundary.

\section{The Knizhnik-Zamolodchikov equation}

We consider the {\small KZ} equation%
\begin{equation}
\triangledown _{j}\Psi =0,\qquad j=1,...,N  \label{kz.1}
\end{equation}%
where the differential operator $\triangledown _{j}$ is defined by use of
Gaudin's Hamiltonians $H_{j}$ (\ref{gm.2}):%
\begin{equation}
\triangledown _{j}=\kappa \frac{\partial }{\partial z_{j}}-H_{j}
\label{m1.29}
\end{equation}%
and $\kappa $ is an arbitrary parameter. The integrable condition for a set
of the {\small KZ}-type differential operators $\triangledown _{j}$,%
\begin{equation}
\left[ \triangledown _{j},\triangledown _{k}\right] =0\qquad for\qquad
j,k=1,..N  \label{kz.2}
\end{equation}%
is satisfied by the condition%
\begin{equation}
\frac{\partial H_{j}}{\partial z_{k}}=\frac{\partial H_{k}}{\partial z_{j}}
\label{kz.3}
\end{equation}

Following the idea of \cite{Babujian1}, we define the hypergeometric
function ${\cal X}(v|z)$ by a set of differential equations%
\begin{eqnarray}
\kappa \frac{\partial {\cal X}(v|z)}{\partial z_{j}} &=&E_{j}{\cal X}%
(v|z),\qquad j=1,...,N  \nonumber \\
\kappa \frac{\partial {\cal X}(v|z)}{\partial v_{a}} &=&f_{a}{\cal X}%
(v|z),\qquad a=1,...,M  \label{kz.4}
\end{eqnarray}%
The integrability of these differential equations follows from the conditions%
\begin{equation}
\frac{\partial E_{j}}{\partial z_{k}}=\frac{\partial E_{k}}{\partial z_{j}}%
,\qquad \frac{\partial E_{j}}{\partial v_{a}}=\frac{\partial f_{a}}{\partial
z_{j}},\qquad \frac{\partial f_{a}}{\partial v_{b}}=\frac{\partial f_{b}}{%
\partial v_{a}}.  \label{kz.5}
\end{equation}%
In fact, it is straightforward to solve the differential equations (\ref%
{kz.4}); its solution is the function%
\begin{eqnarray}
{\cal X}(v|z) &=&\prod\limits_{a}^{M}(\cosh v_{a})^{2/\kappa }[\sinh
(v_{a}-c_{L})\sinh (v_{a}+c_{L})\sinh (v_{a}-c_{R})\sinh
(v_{a}+c_{R})]^{-1/\kappa }  \nonumber \\
&&\times \prod\limits_{j}^{N}(\tanh z_{j})^{1/2\kappa }[\sinh
(z_{j}-c_{L})\sinh (z_{j}+c_{L})\sinh (z_{j}-c_{R})\sinh
(z_{j}+c_{R})]^{1/\kappa }  \nonumber \\
&&\times \prod\limits_{j}^{N}\prod\limits_{a}^{M}[\sinh (z_{j}-v_{a})\sinh
(z_{j}+v_{a})]^{-1/\kappa }\prod\limits_{a<b}^{M}[\sinh (v_{a}-v_{b})\sinh
(v_{a}+v_{b})]^{2/\kappa }  \nonumber \\
&&  \label{kz.6}
\end{eqnarray}

One can introduce the wavefunction $\Psi (z)$ in a integral form, which has
a hypergeometric kernel \cite{Varchenko}, as%
\begin{equation}
\Psi(z)=\oint_{C}\prod\limits_{a}^{M}dv_{a}{\cal X}(v|z)\phi(v|z)
\label{kz.7}
\end{equation}

The integration path $C$ is taken over a closed contour in the Riemann
surface such that the integrand resumes its initial value after $v_{a}$ has
described it. The integral function $\Psi (z)$ is in fact a solution of the 
{\small KZ} equation (\ref{kz.1}).

To prove (\ref{kz.1}) we use the fact that the Bethe state $\phi $ (\ref%
{gm.7}) satisfies%
\begin{equation}
\frac{\partial \phi }{\partial z_{j}}=\sum_{a=1}^{M}\left( \frac{\cosh
(v_{a}-z_{j})}{\sinh ^{2}(v_{a}-z_{j})}-\frac{\cosh (v_{a}+z_{j})}{\sinh
^{2}(v_{a}+z_{j})}\right) \sigma _{j}^{-}\phi _{a}  \label{kz.9}
\end{equation}%
where $\phi _{a}$ is defined in (\ref{gm.8}). One sees that the function $%
\phi _{a}$ does not depend on $v_{a}$. Then equality (\ref{kz.1}) can be
verified as%
\begin{eqnarray}
\kappa \frac{\partial \Psi (z)}{\partial z_{j}} &=&\oint_{C}\left( \kappa 
\frac{\partial {\cal X}}{\partial z_{j}}\phi +\kappa {\cal X}\frac{\partial
\phi }{\partial z_{j}}\right) \prod\limits_{a}^{M}dv_{a}=\oint_{C}\left( 
{\cal X}E_{j}\phi +\kappa {\cal X}\frac{\partial \phi }{\partial z_{j}}%
\right) \prod\limits_{a}^{M}dv_{a}  \nonumber \\
&=&H_{j}\Psi (z)-\oint_{C}\left( \sum_{a=1}^{M}\frac{2\sinh z_{j}\cosh v_{a}%
}{\sinh (z_{j}-v_{a})\sinh (z_{j}+v_{a})}\kappa \frac{\partial {\cal X}}{%
\partial v_{a}}\sigma _{j}^{-}\phi _{a}-\kappa {\cal X}\frac{\partial \phi }{%
\partial z_{j}}\right) \prod\limits_{a}^{M}dv_{a}  \nonumber \\
&=&H_{j}\Psi (z)-\kappa \sum_{a=1}^{M}\oint_{C}\frac{\partial }{\partial
v_{a}}\left( {\cal X}\frac{2\sinh z_{j}\cosh v_{a}}{\sinh (z_{j}-v_{a})\sinh
(z_{j}+v_{a})}\right) dv_{a}\sigma _{j}^{-}\phi _{a}\prod\limits_{b\neq
a}^{M}dv_{b}  \nonumber \\
&=&H_{j}\Psi (z)  \label{kz.10}
\end{eqnarray}

\section{Discussion}

We have constructed and solved a Gaudin magnet with impurity. The integral
representation of the solution for the correspondig {\small KZ} equation was
obtained and its rational limit is here given by

\begin{eqnarray}
\Psi (z)
&=&\oint\limits_{C}dv\prod%
\limits_{a}^{M}[(v_{a}^{2}-c_{L}^{2})(v_{a}^{2}-c_{R}^{2})]^{-1/\kappa
}\prod\limits_{j}^{N}(z_{j})^{1/2\kappa
}[(z_{j}^{2}-c_{L}^{2})(z_{j}^{2}-c_{R}^{2})]^{1/\kappa }  \nonumber \\
&&\times
\prod\limits_{j}^{N}\prod\limits_{a}^{M}[z_{j}^{2}-v_{a}^{2}]^{-1/\kappa
}\prod\limits_{a<b}^{M}[v_{a}^{2}-v_{b}^{2}]^{2/\kappa }  \nonumber \\
&&\times \prod\limits_{a}^{M}\left( \frac{2v_{a}}{v_{a}^{2}-c_{L}^{2}}\sigma
_{L}^{-}+\frac{2v_{a}}{v_{a}^{2}-c_{R}^{2}}\sigma _{R}^{-}+\sum_{j=1}^{N}%
\frac{2v_{a}}{v_{a}^{2}-z_{j}^{2}}\sigma _{j}^{-}\right) \left\vert
0\right\rangle  \label{disc.1}
\end{eqnarray}%
where $dv=\prod\limits_{a}^{M}dv_{a}$.

Now, if we use our Bethe reference state\ (\ref{qism.29}) as a particular
chiral primary field of the defect conformal field theory \cite{DeWolfe}
based in the $su(2)$ {\small WZW} conformal field theory, we conjecture that
the integral representation (\ref{disc.1}) should be the candidate for the
corresponding $M$-point correlation function since it is a solution of the 
{\small KZ} equation \ref{kz.1}. Here we note that the corresponding result
without impurity was already obtained by Hikami \cite{Himaki2}.

In this paper we only considered the $su(2)$ {\small XXZ} spin chain case.
The generalization to the $su(n)$ should be done from the view point of the
Gaudin magnet with impurity \cite{Flume}. Another also interesting case to
be considered is the elliptic {\small XYZ} Gaudin magnet with impurity.

\vspace{0.5cm}

{\bf Acknowledgment:} This work was supported in part by Funda\c{c}\~{a}o de
Amparo \`{a} Pesquisa do Estado de S\~{a}o Paulo-{\small FAPESP}-Brasil and
by Conselho Nacional de Desenvolvimento-{\small CNPq}-Brasil.

\end{document}